\documentclass[a4paper,10pt]{article}
\begin{document}

\title{Speculative Bubbles and Crashes in Stock Markets: 
An Interacting-Agent Model of Speculative Activity} 
\author{Taisei Kaizoji \\ Department of Economics, University of Kiel
\thanks{Olshausenstr. 40, 24118 Kiel, Germany. On leave from Division of Social Sciences, 
International Christian University, Osawa, Mitaka, Tokyo, 181-8585, Japan. E-mail: kaiyoji@icu.ac.jp}}
\date{}
\maketitle

\begin{abstract}
 In this paper we present an interacting-agent model of speculative activity 
explaining bubbles and crashes in stock markets. We describe stock markets through an infinite-range Ising model to formulate the tendency of traders getting influenced by the investment attitude of other traders.
Bubbles and crashes are understood and described qualitatively and quantitatively in terms of the classical phase transitions. 
When the interactions among traders become stronger and reach some critical values, a second-order phase transition and critical behaviour can be observed, and a bull market phase and a bear market phase appear. When the system stays at the bull market phase, speculative bubbles occur in the stock market. For a certain range of the external field that we shall call the investment environment, multistability and hysteresis phenomena are observed. When the investment environment reaches some critical values, the rapid changes in the distribution of investment attitude are caused. The first-order phase transition from a bull market phase to a bear market phase is considered as a stock market crash. \par
Furthermore we estimate the parameters of the model using the actual financial data. As an example of large crashes we analyse Japan crisis (the bubble and the subsequent crash in the Japanese stock market in 1987-1992), and show that the good quality of the fits, as well as the consistency of the values of the parameters are obtained from Japan crisis. 
The results of the empirical study demonstrate that Japan crisis can be explained quite naturally by the model 
that bubbles and crashes have their origin in the collective crowd behaviour of many interacting agents. \\  
{\bf keyword}
Speculative Bubbles; Stock market crash; Phase transition; Mean field approximation; Japan crisis
\end{abstract}

\section{Introduction} 
The booms and the market crashes in financial markets have been an object of study in economics and a history of economy for a long time. Economists \cite{Keynes},  
and Economic historians \cite{Mackay}, \cite{Kindlberger}, \cite{Galbraith} 
have often suggested the importance of psychological factors and irrational factors in explaining historical financial euphoria. As Keynes \cite{Keynes}, a famous 
economist and outstandingly successful investor, acutely pointed out in his book,
{\it The General theory of Employment, Interest and Money}, stock price changes have 
their origin in the collective crowd behaviour of many interacting agents rather than 
fundamental values which can be derived from careful analysis of present conditions 
and future prospects of firms. In a recent paper published in the Economic Journal, 
Lux \cite{Lux1} modelled the idea explicitly and proposed a new theoretical framework to explain 
bubbles and subsequent crashes which links market crashes to the phase 
transitions studied in statistical physics. In his model the emergence of bubbles and crashes is formalised as a self-organising process of infections among heterogeneous 
traders \footnote{For a similar study see also Kaizoji \cite{Kaizoji}.}. In recent independent works, several groups of physicists \cite{SJ1}, \cite{SJ2}, \cite{SJ3}, \cite{JS1}, \cite{JS2}, \cite{JS3}, \cite{F1}, \cite{F2}, \cite{V1}, \cite{V2}, \cite{Gluzman} proposed and demonstrated empirically that large stock market crashes, such as the 1929 and the 1987 crashes, are analogous to critical points. 
They have claimed that financial crashes can be predicted using the idea of 
log-periodic oscillations or by other methods inspired by the physics of critical phenomena \footnote{See also a critical review on this literature \cite{Laloux}.}. 
In this paper we present an interacting-agent model of speculative activity 
explaining bubbles and crashes in stock markets. We describe stock markets through an infinite-range Ising model to formulate the tendency of traders getting influenced by the investment attitude of other traders. Bubbles and crashes are understood and described qualitatively and quantitatively in terms of the classical phase transitions \footnote{A similar idea has been developed in the Cont-Bouchaud model with an Ising modification \cite{Chowdhury} from another point of view. For a related study see also \cite{Holyst}. They study phase transitions in the social Ising models of opinion formation.}. Although the interacting-agent hypothesis \cite{Lux2} is advocated as an alternative approach to the efficient market hypothesis (or rational expectation hypothesis) \cite{Fama}, little attention has been given to the point how probabilistic rules, that agents switch their investment attitude, are connected with their decision-making or their expectation formations. Our interacting-agent model follows the line of Lux \cite{Lux1}, but differs from his work in the respect that we model speculative activity here from a viewpoint of traders' decision-making. 
The decision-making of interacting-agents will be formalised by {\it minimum energy 
principle}, and the stationary probability distribution on traders' investment 
attitudes will be derived. Next, the stationary states of the system and speculative 
dynamics are analysed by using the mean field approximation. It is suggested that the mean field approximation can be considered as a mathematical formularisation of {\it Keynes' beauty contest}.  
There are three basic stationary states in the system: a bull market equilibrium, a bear market equilibrium, and a fundamental equilibrium. We show that the variation of parameters like the bandwagon effect or the investment environment, which corresponds to the external field, can change the size of cluster of traders' investment attitude or make the system jump to another market phase. When the bandwagon effect reaches some critical value, a second-order phase transition and critical behaviour can be observed. There is a symmetry breaking at the fundamental equilibrium, and two stable equilibria, the bull market equilibrium and bear market equilibrium appear. When the system stays in the bull market equilibrium, speculative bubble occurs in the stock market. For a certain range of the investment environment multistability and hysteresis phenomena are observed. 
When the investment environment reaches some critical values, the rapid changes (the first-order phase transitions) in the distribution of investment attitude are caused. The phase transition from a bull 
market phase to a bear market phase is considered as a stock market crash. \par  
Then, we estimate the parameters of the interacting-agent model using an 
actual financial data. As an example of large crashes we will analyse the Japan crisis (bubble and crash in Japanese stock market in 1987-1992). The estimated equation attempts to explain Japan crisis over 6 year period 1987-1992, and was constructed using monthly adjusted data for the first difference of TOPIX and the investment environment which is defined below. Results of estimation suggest that the traders in the Japanese stock market stayed the bull market equilibria, so that the speculative bubbles were caused by the strong bandwagon effect and betterment of the investment environment in 3 year period 1987-1989, but a turn for the worse 
of the investment environment in 1990 gave cause to the first-order phase transition from a bull market phase to a bear market phase. We will demonstrate that the 
market-phase transition occurred in March 1990.  
In Section 2 we construct the model. In Section 3 we investigate the relationship 
between crashes and the phase transitions. We implement an empirical study of Japan 
crisis in Section 4. We give some concluding remarks in Section 5. 
\section{An Interacting-Agent Model of Speculative Activity} 
We think of the stock market that large numbers of traders participate in trading. The stock market consists of N traders (members of a trader group). Traders are indexed by $ j = 1, 2, ........, N $. We assume that each of them can share one of two investment attitudes, buyer or seller, and buy or sell a fixed amount of stock (q) in a period.  $ x_{i} $ denotes the investment attitude of trader $ i $ at a period. The investment attitude $ x_{i} $ is defined as follows: if trader $ i $ is the buyer of the stock at a period, then $ x_{i} = + 1 $. If trader $ i $, in contrast, is the seller of the stock at a period, then $ x_{i} = - 1 $. 
\subsection{Decision-Making of traders} 
In the stock market the price changes are subject to the law of demand and supply, that the price rises when there is excess demand, and the price falls when there is excess supply. It seems natural to assume that the price raises if the number of the buyer exceeds the number of the seller because there may be excess demand, and the price falls if the number of seller exceeds the number of the seller because there may be 
excess supply. Thus a trader, who expects a certain exchange profit through trading, will predict every other traders' behaviour, and will choose the same behaviour as the other traders' behaviour as thoroughly as possible he could. 
The decision-making of traders will be also influenced by changes of the firm's fundamental value, which can be derived from analysis of present conditions and future prospects of the firm, and the return on the alternative asset (e.g. bonds). 
For simplicity of an empirical analysis we will use the ratio of ordinary profits to total capital that is a typical measure of investment, as a proxy for changes of the fundamental value, and the long-term interest rate as a proxy for changes of the return on the alternative asset. Furthermore we define the investment environment as \par
{\it investment environment = ratio of ordinary profits to total capital 
- long-term interest rate}. \par
When the investment environment increases (decreases) a trader may think that 
now is the time for him to buy (sell) the stock. 
Formally let us assume that the investment attitude of trader $ i $ is determined by minimisation of the following {\it disagreement} function $ e_i(x) $, 
\begin{equation}
    e_i(x) = - \frac{1}{2} \sum^N_{j=1} a_{ij} x_i x_j - b_i s x_i. 
   \label{eqn:a1}
   \end{equation}
where $ a_{ij} $ denotes the strength of trader $ j $'s influence on trader $ i $, and $ b_i $ denotes the strength of the reaction of trader $ i $ upon the change of the investment environment $ s $ which may be interpreted as an external field, and $ x $ denotes the vector of investment attitude $ x = (x_1, x_2,......x_N) $.  
The optimisation problem that should be solved for every trader to achieve minimisation of their disagreement functions $ e_i(x) $ at the same time is formalised by 
   \begin{equation}
   \min E(x) = - \frac{1}{2} \sum^N_{i=1} \sum^N_{j=1} a_{ij} x_i x_j 
               - \sum^N_{i=1} b_i s x_i. 
   \label{eqn:a2}
   \end{equation}
Now let us assume that trader's decision making is subject to a probabilistic rule. 
The summation over all possible configurations of agents' investment attitude $ x = (x_1,.....,x_N) $ is computationally explosive with size of the number of trader $ N $.Therefore under the circumstance that a large number of traders participates into trading, a probabilistic setting may be one of best means to analyse the collective behaviour of the many interacting traders. Let us introduce a random 
variable $ x^k = (x^k_1, x^k_2,......, x^k_N) $, $ k = 1, 2,....., K $. The state of the agents' investment attitude $ x^k $ occur with probability $ P(x^k) = \mbox{Prob}(x^k) $ 
with the requirement $ 0 < P(x^k) < 1 $ and $ \sum^K_{k=1} P(x^k) = 1 $. We define the amount of uncertainty before the occurrence of the state $ x^k $ with probability $ P(x^k) $ as the logarithmic function: $ I(x^k) = - \log P(x^k) $. 
Under these assumptions the above optimisation problem is formalised by 
    \begin{equation}
    \min \langle E(x) \rangle = \sum^N_{k=1} P(x^k) E(x^k) 
    \label{eqn:a3}
    \end{equation}
subject to 
$ H = - \sum^N_{k=1} P(x^k) \log P(x^k), \quad \sum^N_{k=-N} P(x^k) = 1 $, \par
where $ E(x^k) = \frac{1}{2} \sum^N_{i=1} E_i(x^k) $. 
$ x^k $ is a state, and $ H $ is information entropy. $ P(x^k) $ is the relative frequency the occurrence of the state $ x^k $. The well-known solutions of the above optimisation problem is 
   \begin{equation}
   P(x^k) = \frac{1}{Z} \exp(- \mu E(x^k)), \quad 
   Z = \sum^K_{k=1} \exp(- \mu E(x^k)) \quad k = 1, 2,....., K. 
   \label{eqn:a4}
   \end{equation}
where the parameter $ \mu $ may be interested as a {\it market temperature} describing a degree of randomness in the behaviour of traders . The probability distribution $ P(x^k) $ is called the {\it Boltzmann distribution} where $ P(x^k) $ is the probability that the traders' investment attitude is in the state $ k $ with the function $ E(x^k) $, and $ Z $ is the partition function. We call the optimising behaviour of the traders with interaction among the other traders a {\it relative expectation formation}. 
\subsection{The volume of investment}
The trading volume should depends upon the investment attitudes of all traders. Since traders are supposed to either buy or sell a fixed amount of stock (q) in a period, the aggregate excess demand for stock at a period is given 
by $ q \sum^N_{i=1} x_i $. 
\subsection{The price adjustment processes}
We assume the existence of a market-maker whose function is to adjust the price. 
If the excess demand $ q x $ is positive (negative), the market maker raises (reduces) the stock price for the next period. Precisely, the new price is calculated as the previous price $ y_t $ plus some fraction of the excess demand of the previous period according: $ \Delta y_{t+1}  = \lambda q \sum^N_{i=1} x_{it} $, 
where $ x_{it} $ denotes the investment attitude of trader $ i $ at period $ t $, $ \Delta y_{t+1} $ the price change from the current period to the next period, i.e. $ \Delta y_{t+1} = y_{t+1} - y_{t} $, and the parameter $ \lambda $ represents the speed of adjustment of the market price. At the equilibrium prices that clear the market there should exist an equal number of buyers or sellers, i.e.: 
$  \sum^N_{i=1} x_{it} = 0 $. Using the Boltzmann distribution (\ref{eqn:a4}) the mean value of the price changes $ \Delta \bar{y} $ is given by 
$ \Delta \bar{y} = \sum^N_{k=1} P(x^k) q \sum^N_{i=1} x^k_i $. 
\subsection{Mean-Field Approximation : Keynes' Beauty Contest} 
We can derive analytically the stationary states of the traders' investment attitude using a mean field approximation which is a well known technique in statistical physics. 
 Let us replace the discrete summation of the investment attitude with the mean field variable, $ \langle x \rangle = \langle (1/N) \sum^N_{i=1} x_i \rangle $. 
We additionally assume that the parameter $ a_{ij} $ is equal to $ a $ $ a_{ij} = a $ and $ b_i = b $ for every trader. In this case our interacting-agent model reduces to the Ising model with long-range interactions. Then the function (\ref{eqn:a2}) is approximated by $  E(x) \approx - (1/2) \sum^N_{i=1} \sum^N_{j=1} 
a N \langle x \rangle x_i - \sum^N_{i=1} b s x_i  $. Since the mean value of the stationary distribution is given by 
$ \langle E(x) \rangle = - \partial \log Z /\partial \mu $,  
the mean field $ \langle x \rangle $ is 
   \begin{equation}
     \langle x \rangle = \tanh(\mu a N \langle x \rangle + \mu b s). 
   \label{eqn:a5} 
   \end{equation}  
The first term of the right side of Eq. (\ref{eqn:a5}) represents that traders 
tend to adopt the same investment attitude as prediction of the average investment 
attitude and the second term represents the influence of the change of investment 
environment to traders' investment attitude. The first term may be interpreted as 
a mathematical formularisation of {\it Keynes' beauty contest}. 
Keynes \cite{Keynes} argued that stock prices are not only determined by the 
firm's fundamental value, but in addition mass psychology and investors expectations 
influence financial markets significantly. It was his opinion that professional 
investors prefer to devote their energies not to estimating fundamental values, but 
rather to analysing how the crowd of investors is likely to behave in the future. 
As a result, he said, most persons are {\it largely concerned, not with making superior long-term forecasts of 
the probable yield of an investment over its whole life, but with foreseeing changes 
in the conventional basis of valuation a short time ahead of the general public}. 
Keynes used his famous {\it beauty contest} as a parable to stock markets. 
In order to predict the winner of a beauty contest, objective beauty is not much 
important, but knowledge or prediction of others' predictions of beauty is much more 
relevant. In Keynes view, the optimal strategy is not to pick those faces the player 
thinks prettiest, but those the other players are likely to be about what the 
average opinion will be, or to proceed even further along this sequence. \par
\subsection{Speculative Price Dynamics}
We assume that the traders' investment attitude changes simultaneously (synchronous 
dynamics) in discrete time steps. We represent the dynamics of the investment attitude 
using a straightforward iteration of Equation (\ref{eqn:a5}), such that:  
  \begin{equation} 
  \langle x \rangle_{t+1} = \tanh(\alpha \langle x \rangle_t 
  + \beta s_t), \quad \alpha = \mu a N, \beta = \mu b. 
  \label{eqn:a6}
  \end{equation}
We call $ \alpha $ the bandwagon coefficient and $ \beta $ the investment environment coefficient below because $ \alpha $ may be interpreted as a parameter that denotes the strength that traders chase the price trend. 
 Next, let us approximate the adjustment process of the price using the mean field 
$ \langle x \rangle_t $. 
We can get the following speculative price dynamics, 
  \begin{equation}
   \langle \Delta y \rangle_{t+1} = \lambda N \langle x \rangle_t. 
   \label{eqn:a7}
  \end{equation}
Inserting the mean field of the price change, $ \langle \Delta y \rangle_{t} $ 
into eq. \ref{eqn:a6}, the dynamics of the mean field of 
$ \langle x \rangle_t $ can be rewritten as $ \langle x \rangle_{t+1} = \tanh(\frac{\mu a}{\lambda} \langle \Delta y \rangle_{t+1} + \mu b s_t) $. This equation shows that the traders result in basing their trading decisions on an analysis of the price trend $ \langle \Delta y \rangle_{t+1} $ as well as the change of the investment environment $ s_t $. 
\section{Speculative Bubbles and Crashes} 
The stationary states of the mean field $ \langle x \rangle $ satisfy the following 
\begin{equation}
  \langle x \rangle_t = \tanh (\alpha \langle x \rangle_t + 
  \beta s) \equiv f(\langle x \rangle_t , s). 
    \label{eqn:a8}
  \end{equation} 
Eq. \ref{eqn:a8} has a unique solution $ \langle x \rangle $ for arbitrary $ \beta s $, when $ \alpha $ is less than $ 1 $. The stationary state $ \langle x 
\rangle^*   $ with $ \beta s = 0 $ is called as the {\it fundamental equilibrium} which is stable 
and corresponds to the maximum of the stationary distribution of the investment 
attitude $ P(x^*) $. In this stable equilibrium there is equal numbers of traders 
sharing both investment attitudes on average. \par
Moving in the parameter space $ (\alpha , \beta) $ and starting from different 
configurations one have several possible scenarios of market-phase transitions. 
To begin with, let us consider the case that the investment environment $ s $ is equal to zero. In theory, when the bandwagon coefficient $ \alpha $ is increased 
starting from the phase at $ 0 < \alpha < 1 $, there is a symmetry breaking at critical 
point, $ \alpha = 1 $, and two different market phases appear, that is, the {\it bear market} and the {\it bull market}. In analogy to physical systems the transitions may be called the second-order phase transition. 
Fig. 1 illustrates the two graphical solutions of the Eq. (\ref{eqn:a8}) 
for $ \alpha > 1 $ and $ s = 0 $ and for $ 0 < \alpha < 1 $ and $ s = 0 $. The figure 
shows that for $ 0 < \alpha < 1 $ and $ s = 0 $ the fundamental equilibrium is unique 
and stable, but for $ \alpha > 1 $ and $ s = 0 $ the fundamental equilibrium $ 0 $ becomes 
unstable, and the two new equilibria, the bull market equilibrium $ a $ and the bear 
market equilibrium $ b $ are stable. At the bull market equilibrium more than half 
number of traders is buyer, so that the speculative bubbles occur in the stock market. 
Then, let us consider the effect of changes of the investment environment $ s $ in 
the case with a weak bandwagon coefficient, $ 0 < \alpha < 1 $ and a positive investment 
environment coefficient, $ \beta > 0 $. Fig. 2 shows that as the investment 
environment $ s $ changes for the better ($ s > 0 $), the system shifts to the 
bull market phase. 
By contrast, as the investment environment $ s $ changes for the worse ($ s < 0 $),  
the system shifts to the bear market phase. Therefore when the bandwagon effect is 
weak, the stock price go up or down slowly according rises or falls in the investment 
environment. \par
Finally, let us consider the effect of changes of $ \beta s $ in the case with a 
strong bandwagon coefficient $ \alpha > 1 $. In this case multistability and 
hysteresis phenomena in the distribution of investment attitude, as well as 
market-phase transitions are observed. 
The system has three equilibria, when $ \alpha > 1 $ and $ |s| < s^* $, where $ s^* 
$ is determined by the equation, 
\begin{equation} 
  \cosh^2 [\beta s \pm \sqrt{\alpha (\alpha - 1)}] = \alpha. 
  \label{eqn:a9}
\end{equation}
Two maximum of the stationary distribution on the investment attitude 
are found at $ \langle x \rangle^- $ and $ \langle x \rangle^+ $ and one minimum 
at $ \langle x \rangle^* $. 
 For $ |s| = s^* $ and $ \alpha > 1 $, two of the three equilibria coincide at 
\begin{equation}
  \langle x \rangle_c = \sqrt{(\alpha - 1)/\alpha}. 
  \label{eqn:a10}
\end{equation}
For $ |s| > s^* $ and $ \alpha > 1 $ the two of three equilibria vanishes and only 
one equilibrium remains. In analogy to physical systems the transitions may be called 
first-order phase transitions. 
Fig. 3 illustrates the effect of changes in the investment 
environment in 
the case with a strong bandwagon effect $ \alpha > 1 $. In this case  
speculative bubbles and market crashes occur. 
When the investment environment changes for the 
better (worse), the curve that denotes $ f(\langle x \rangle_t , s) $, shifts upward 
(downward). As the investment environment $ s $ keeps on rising, and when it reach 
a critical value, that is, $ |s| = s^* $, the bear market equilibrium and the 
fundamental equilibrium vanish, and the bull market equilibrium becomes an unique 
equilibrium of the system, so that the speculative bubble occurs in the stock market. 
Even though the investment environment changes for the worse, the hysteresis 
phenomena are observed in a range of the investment environment that can be calculated 
by solving Eq. (\ref{eqn:a9}). In other words the speculative bubble continues in a range of the investment environment. However when the negative impact of the investment environment reaches a critical value, that is, $ |s| = s^* $, the bull market vanish. Further decrease of $ s $ cause 
the market-phase transition from a bull market to a bear market. In a bear market 
more than half number of trader is seller on average, so that the stock price continues 
to fall on average. Thus this market phase transition may be considered as the stock market crash.
\section{Empirical Analysis : Japan Crisis}
Perhaps the most spectacular boom and bust of the late twentieth century 
involved the Japan's stock markets in 1987-1992. Stock prices increased from 
1982 to 1989. At their peak in December 1989, Japanese stocks had a total market value 
of about 4 trillion, almost 1.5 times the value of all U.S. equities and close to 
45 percent of the world's equity market capitalisation. The fall was almost as extreme 
as the U.S. stock-market crash from the end of 1929 to mid-1932. The Japanese (Nikkei) 
stock-market index reached a high of almost 40,000 yen in the end of 1989. By 
mid-August 1992, the index had declined to 14,309 yen, a drop of about 
63 percent. \par
In this section we will estimate the parameter vector $ (\alpha , \beta) $ for the 
Japan's stock market in 1987-1992. We use monthly adjusted data for the first 
difference of Tokyo Stock Price Index (TOPIX) and the investment environment over 
the period 1988 to 1992. In order to get the mean value of the stock price changes,  
TOPIX is adjusted through the use of a simple centred 12 point moving average, and 
are normalised into the range of $ -1 $ to $ +1 $ by dividing by the maximum value 
of the absolute value of the price changes. The normalised price change is defined 
as $ \langle \Delta p \rangle_t $.  
Fig. 4 and Fig. 5 show the normalised stock price changes and the 
investment environment respectively. This normalised stock price change $ \langle \Delta p \rangle_t $ is used as a proxy of the mean field $ \langle x \rangle_t $. 
Inserting $ \langle \Delta p \rangle_t $ for Eq. (\ref{eqn:a8}), we get 
$  f(\langle \Delta p \rangle_t , s_t) 
   = \tanh(\alpha \langle \Delta p \rangle_t + \beta s_t) $. 
\subsection{Model estimation: The gradient-descent algorithm}
The bandwagon coefficient $ \alpha $ and the investment environment coefficient 
$ \beta $ should be estimated using any one of estimation 
technique. Since we cannot get the analytical solution because the function $ f(\langle \Delta p \rangle_t , s_t) $ is non-linear, we use the {\it gradient-descent 
algorithm} for the parameter estimation \footnote{The gradient-descent method is 
often used for training multilayer feedforward networks. It is called the back-
propagation learning algorithm which is one of the most important historical 
developments in neural networks [Rumelhart et al. \cite{Rumelhart}.}. 
The gradient-descent algorithm is a stable and robust procedure for minimising the 
following one-step-prediction error function 
\begin{equation}
  E(\alpha_k, \beta_k) 
  = \frac{1}{2} \sum^n_{t=1} [\langle \Delta p \rangle_{t} - 
  f(\langle \Delta p \rangle_{t-1}, s_{t-1})]^2.  
   \label{eqn:a33}
\end{equation}
More specifically, the gradient-descent algorithm changes the parameter vector 
$ (\alpha_k , \beta_k) $ to satisfy the following condition: 
\begin{equation}
   \Delta E(\alpha_k , \beta_k) = \frac{\partial E(\alpha_k, \beta_k)}
 {\partial \alpha_k} \Delta \alpha_k + \frac{\partial E(\alpha_k, \beta_k)}
 {\partial \beta_k} \Delta \beta_k < 0, 
   \label{eqn:a34}
\end{equation}
where $ \Delta E(\alpha_k , \beta_k) = E(\alpha_{k} , \beta_{k}) - 
E(\alpha_{k-1} , \beta_{k-1}) $, $ \Delta \alpha_k = 
\alpha_{k} - \alpha_{k-1} $, 
and $ \Delta \beta_k = \beta_{k} - \beta_{k-1} $. 
To accomplish this the gradient-descent algorithm adjusts each parameter $ \alpha_k 
$ and $ \beta_k $ by amounts $ \Delta \alpha_{k} $ and $ \Delta \beta_{k} $ proportional 
to the negative of the gradient of $ E(\alpha_k , \beta_k) $ at the current location: 
\begin{equation}
 \alpha_{k+1} = \alpha_k - \eta \frac{\partial E(\alpha_k, \beta_k)}
 {\partial \alpha_k}, \quad 
 \beta_{k+1} = \beta_k - \eta \frac{\partial E(\alpha_k, \beta_k)}
 {\partial \beta_k} 
   \label{eqn:a35}
\end{equation}
where $ \eta $ is a learning rate. The gradient-descent rule necessarily decreases the error with a small value of 
$ \eta $. If the error function (\ref{eqn:a33}), thus, has a single minimum 
at $ E(\alpha_k , \beta_k) = 0 $, then the parameters $ (\alpha_k , \beta_k) $ 
approaches the optimal values with enough iterations. The error function 
$ f(\alpha_k , \beta_k) $, however, is non-linear, and hence it is possible that 
the error function (\ref{eqn:a33}) may have {\it local minima} besides the global 
minimum at $ E(\alpha_k , \beta_k) = 0 $. In this case the gradient-descent rule may 
become stuck at such a local minimum. 
To check convergence of the gradient-descent method to the global minimum we estimate 
the parameters for a variety of alternative start up parameters 
$ (\alpha_0 , \beta_0) $ with $ \eta = 0.01 $. As a consequence the gradient-descent 
 rule estimated the same values of the parameters, 
$ (\alpha^* , \beta^*) = (1.04 , 0.5) $ with enough iterations. 
\subsection{Results}
The fit of the estimated equation can be seen graphically in Fig. (6), 
which compare the actual and forecasted series over one period, respectively. The correlation coefficient that indicates goodness of fit is $ 0.994 $. The capacity of the model to forecast adequately in comparison with competing models is an important element in the evaluation of its 
overall performance. Let us consider the linear regression model 
$ \langle \Delta{y} \rangle_{t+1} =  c \langle \Delta{y} \rangle_t + d s_t $. 
The regression results are $ \langle \Delta{p} \rangle_{t+1} = 0.855 \langle 
 \Delta{p} \rangle_t + 3.81 s_t $. The correlation coefficient of the linear regression model is equal to 0.93, and is lower that that of the interacting-agent model. Root mean squared errors (RMSEs) of the interacting-agent model is equal to 0.05, and that of the linear regression model is equal to 0.148 over the period 1987-1992. The interacting-agent model has a smaller RMSE than that of the linear regression model. By the traditional RMSE criterion the interacting-agent model is, thus, superior to the linear regression model in term of their forecasting performance. In conclusion we can say that the model not only explains the Japan crisis but also correctly forecasts changing the stock price over the period 1987-1992. 
Given that the set of parameters ($ \alpha , \beta $) is equal to (1.04, 0.5), 
the critical point of the investment environment at which the phase transition are caused from the bull market to bear market are calculated from the theoretical results 
in the preceding section. One can say that the phase transition from the bull market 
to the bear market occurs when the investment environment is below $ -0.007 $ under $ (\alpha , \beta) = (1.04, 0.5) $. By contrast the phase transition from the bear 
market to the bull market occurs when the investment environment is beyond $ 0.007 $ under $ (\alpha , \beta) = (1.04, 0.5) $. 
\begin{table}
  \begin{center}
  \begin{tabular}{c|ccc} \hline
  Period & $ \langle \Delta p \rangle $ & 
  Predicted value & $ s_t $ \\ \hline  
 1989.09  & 0.4591 & 0.4487   & 0.0182 \\
 1989.10  & 0.4468 & 0.4361   & 0.0122 \\
 1989.11  & 0.4531 & 0.4408   & 0.011 \\
 1989.12  & 0.4369 & 0.4258   & 0.0075 \\
 1990.01  & 0.2272 & 0.2294   & -0.002 \\
 1990.02  & 0.0985 & 0.0986   & -0.0055 \\ \hline
 1990.03 & -0.2013 & -0.2097  & -0.0099 \\ \hline
 1990.04 & -0.2354 & -0.2415  & -0.0066 \\ 
 1990.05 & -0.0845 & -0.0871  & -0.0004 \\
 1990.06 & -0.0883 & -0.0924  & -0.0029 \\
 1990.07 & -0.3136 & -0.3159  & -0.0067 \\
 1990.08 & -0.5245 & -0.4992  & -0.0136 \\ \hline
  \end{tabular}
  \caption{The stock market crash in the Japanese stock market}
\label{tab1}
\end{center}
\end{table}
From theoretical view point one may say that bursting speculative bubbles will begin 
when the investment environment is below $ -0.007 $, and the process of the collapse of the stock market continues till the investment environment 
is beyond $ 0.007 $. \par 
In the real world the investment environment became below $ - 0.007 $ at March 1990 for the first time and continued to be negative values since then over the period 
1987-1992. The fall of the actual price in the Japan's stock market began at March 1990, and the stock prices continued to fall over the period 1990-1992. 
On these grounds we have come to the conclusion that theory and practice are in a perfect harmony. (See Table \ref{tab1}.)  
\section{Conclusion}
This paper presents an interacting-agent model of speculative activity explaining 
the bubbles and crashes in stock markets in terms of a mean field approximation. 
We show theoretically that these phenomena are easy to understand as the market-
phase transitions. Bubbles and crashes are corresponding to the second-order 
transitions and the first-order transitions respectively.  
Furthermore we estimate the parameters of the model for the Japanese stock market in 1987-1992. The empirical results demonstrate that theory and practice are in a perfect harmony. This fact justifies our model that bubbles and crashes have their origin in the collective crowd behaviour of many interacting agents. 
\section{Acknowledgements} 
I would like to thank Masao Mori and Thomas Lux for helpful comments and suggestions to an earlier version of this paper.

\section{Figure Captions} 
Figure 1. The second-order phase transition. Numerical solution of Eq.(\ref{eqn:a8}) for 
$ s = 0 $. The curve is the RHS of \ref{eqn:a8} plotted for different 
values of $ \alpha $: the AA line - $ \alpha = 0.9 $, the BB line - 
$ \alpha = 2 $.\\

Figure 2. Numerical solution of Eq.(\ref{eqn:a8}) for $ \alpha = 0.8 $ and $ \beta = 1 $. $ f(\langle x \rangle_t , s) $ is the RHS of (8) plotted for different values of $ \alpha $: the AA line - $ s = 0.3 $, the BB line - $ s = -0.3 $.\\
 
Figure 3. The first-order transition: Speculative bubble and crash. 
Numerical solution of Eq. (\ref{eqn:a8}) for $ \alpha = 1.8 $ and $ \beta = 1 $. 
The curves is $ f(\langle x \rangle_t , s) $ plotted for different values of the 
investment environment: the AA line - $ s = 0.41 $, the BB line - $ s = -0.41 $.\\

Figure 4. The adjusted stock price change, $ \langle \Delta p \rangle $: January 1987 - December 1992. \\
 
Figure 5. The Japan's investment environment $ s_t $: January 1987 - December 1992.\\

Figure 6. One-step-forecast of the stock price changes: 
January 1987 - December 1992.
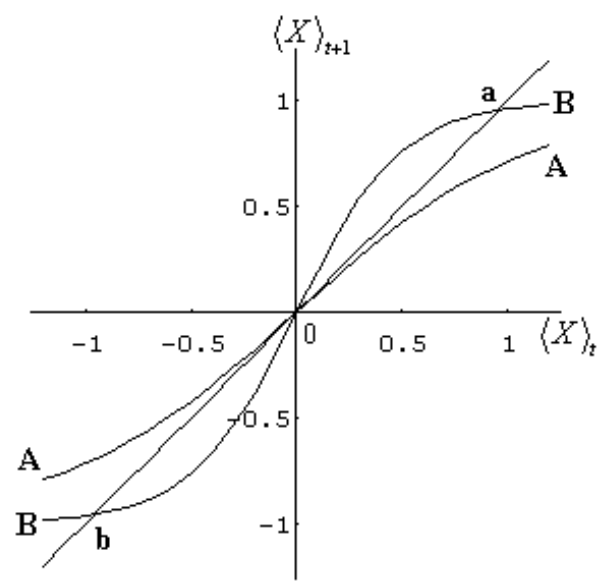
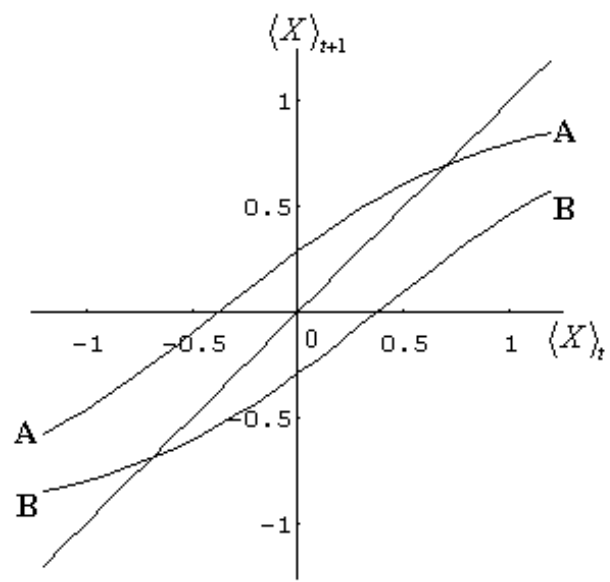
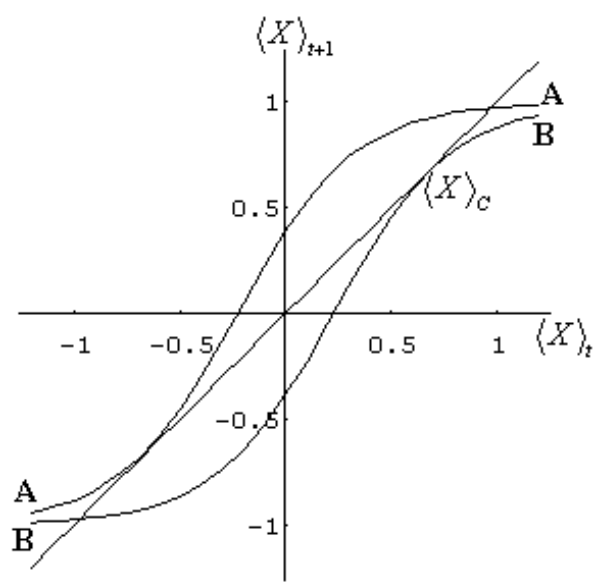
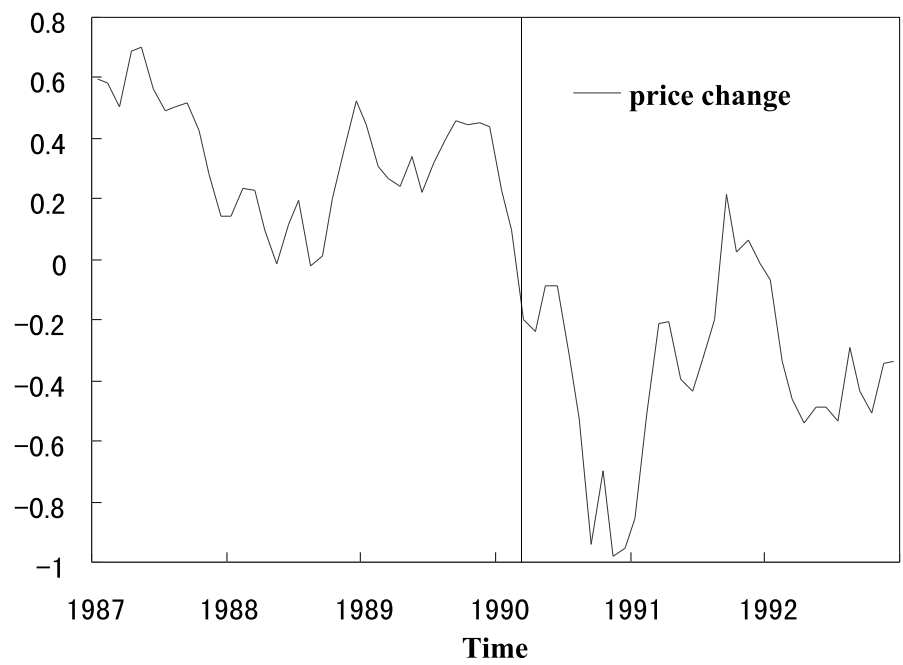
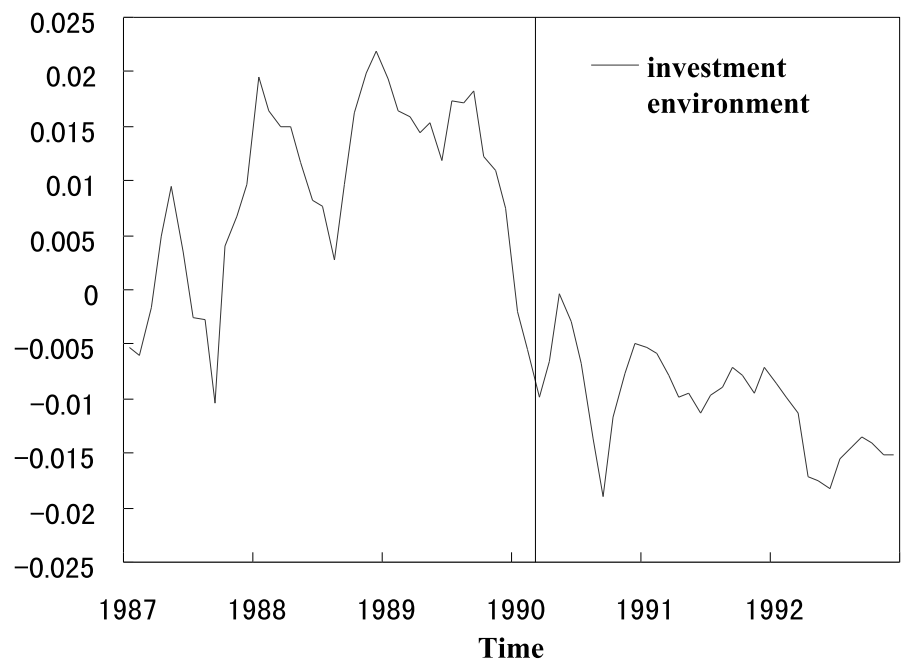
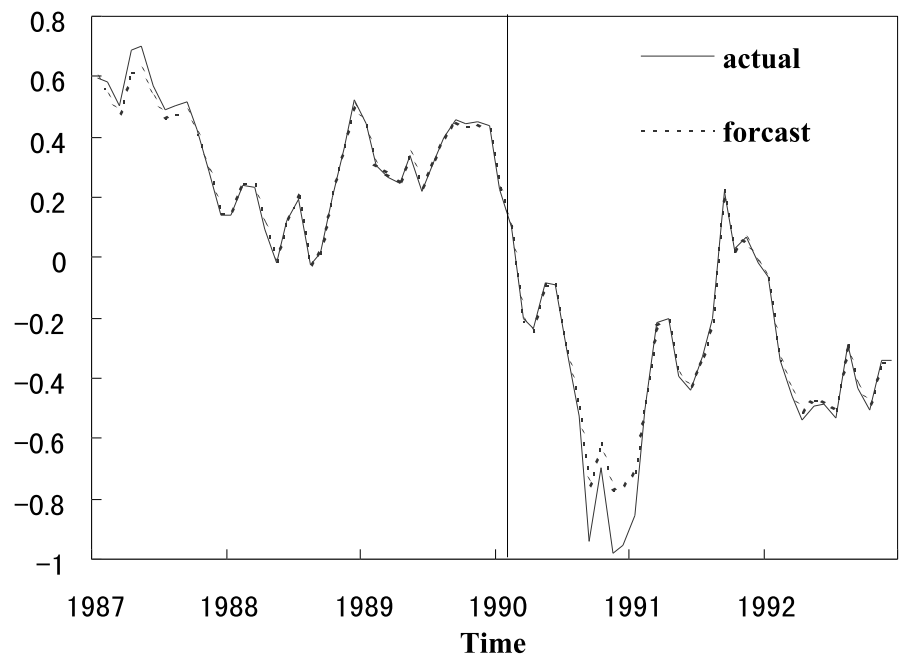
\end{document}